\documentclass[%
 reprint,
showpacs,
 amsmath,
 amssymb,
 aps,
 prl,
floatfix,
]{revtex4-1}

\usepackage{graphicx}
\usepackage{dcolumn}
\usepackage{bm}

\begin{document}


\title{Subcritical laminar-turbulence transition with wide domains in simple two-dimensional Navier-Stokes flow without walls}
\author{Yoshiki Hiruta}
\email[]{hiruta@kyoryu.scphys.kyoto-u.ac.jp}
\author{Sadayoshi Toh}
\affiliation{%
  Division of Physics and Astronomy, Graduate School of Science, Kyoto University, Japan}%

\date{\today}
\begin{abstract}
 
 We have confirmed numerically that a subcritical laminar-turbulence transition
 that belongs to directed percolation (DP) universality class occurs 
 in a purely two-dimensional (2D) simple Navier-Stokes (NS) flow
 without any walls.
 The flow is called (extended) Kolmogorov flow which is governed by 2D
 NS equation in a doubly periodic box with  a linear drag and
 a finite flow rate in the direction in which Galilean invariance is broken.
 We examine the mechanism of DP class transition
 focusing on the role of the additional control parameters:
 the drag coefficient and the flow rate.
 The drag kills coherent active structures of the system size.
 The flow rate interferes the growth of weak disturbances.
 These two effects control two essential and intrinsic elements
 of an absorbing state phase transition, i.e., the existence of an
 absorbing state and the locality of active dynamical structures.
 We also discuss what physically corresponds to the additional parameters
  in general flow. 
 
\end{abstract}

\maketitle

Turbulence, macroscopically complex behavior,
is seen in a wide range of phenomena in nature
not only in a scale of our daily life such as a streamflow of a river
and experiments in laboratory\cite{Reynolds1883a,Coles1965}
but also 
from  quantum turbulence of superfluid\cite{Kobayashi2005,Tsubota2017}
to cosmic scales\cite{Bizon2011}.
It can be regarded as a class of solutions of hydrodynamic equations
whose energy is distributed into a wide range of scales.
However, it is not completely uncovered for a long time
how to sustain turbulence even for 
a flow in a circular pipe driven by pressure gradient\cite{Reynolds1883a,Eckhardt2007,Willis2008}.

At the middle of 20 century, the scenarios of supercritical transition 
have been submitted from the point of bifurcation of dynamical systems\cite{Landau1944,Hopf1948,Ruelle1971}.
In a phase space a fixed point corresponding
to a  laminar flow becomes linearly unstable at a finite critical non-dimensional velocity,
called Reynolds number in general,
and a cascade of bifurcations  leads chaotic behavior corresponding to turbulence.
This type of scenario explains a part of turbulence transition
well such as thermal convection
and wake behind cylinder\cite{Lorenz1963}.

However, linear stability theory is unable to explain subcritical transition to turbulence.
It is well known that representative canonical wall-bounded flows
such as pipe flow and Couette flow have sustained turbulent solutions 
starting from a finite amplitude of disturbance even when the laminar flow is linearly stable.
Such subcritical transition is  difficult to be elucidated 
since dominant modes are unclear because of nonlinearity.

At the end of 20th century, a conjecture was submitted 
to connect (subcritical) laminar-turbulence transition 
with nonequilibrium phase transition common in statistical physics\cite{Pomeau1986}.
This idea is that turbulence transition would belong to Directed percolation(DP) universality class
if there exist absorbing laminar state and turbulent regions which
expand and decay stochastically and locally.
Universality class tells us approximate predictions of spatio-temporal structure of 
disordered regions at least\cite{Allhoff2012}.

Recent developments of understandings of spatially-localized turbulent states (SLT)
enable us to identify candidates for physical correspondence to a unit of turbulence 
such as puff in pipe flow\cite{Avila2011}.
Under detailed laboratory and numerically experimental investigations of subcritical turbulence transition,
critical exponents for turbulence fraction and spatio-temporal structure
are satisfactorily consistent with those of  DP universality class 
for sustained wall turbulence\cite{Lemoult2016}, channel flow turbulence\cite{Sano2015}
and other flow\cite{Chantry2017a}.

DP universality class is supposed to appear under simple conjectures
like locality and stochastic behavior of dynamics\cite{Hinrichsen2000a}.
So, DP behavior arises for a wide range of models of epidemics and synchronization,
experiments for turbulent liquid crystals\cite{Takeuchi2007,Takeuchi2009}.
However, from the point of governing equations of fluid dynamics,
which are nonlinear partial differential equations,
the dynamics of fluid should strongly depend on way to drive, boundary conditions and spatial dimension.
Thus, how to emerge and sustain SLT
and locality of dynamics are not obvious and remain open-problem\cite{Kawahara2012}.
Generally, flows with material wall boundaries  are difficult to be studied
since the walls induce both explicit and implicit inhomogeneities in  the flows.
This difficulty prevent us from relating  large scale behavior of
turbulent system such as subcritical regime of turbulence to relatively
well-known dynamics of small systems\cite{Nagata1990,Jimenez1990,Waleffe1997,Avila2013,Willis2013,Kreilos2012a,Mellibovsky2012,Kawahara2001,Itano2001,Kawahara2012,GIBSON2008}.
Thus, a tractable Navier-Stokes model to investigate subcritical
transition is promised.

One of the most simple two-dimensional (2D) flows sustaining
turbulence called Kolmogorov flow was submitted 
to investigate routes to turbulence\cite{Arnold1960b,Meshalkin1961,Sivashinsky1985,Gotoh1983,Yamada1986a,Marchioro1987}.
Kolmogorov flow is used to obtain many characteristics of
solutions of Navier-Stokes equation and as a test field of novel idea of
understanding complex solutions\cite{Gallet2013,Chandler2013,Farazmand2015,Lucas2015a}.
Recently, solutions in which spatially-localized chaotic regions and steady regions
coexist were found even in 2D Kolmogorov flow\cite{Lucas2014c,Hiruta2015,Hiruta2017a}.

\begin{figure}[h]
  \includegraphics[width=0.49\linewidth]{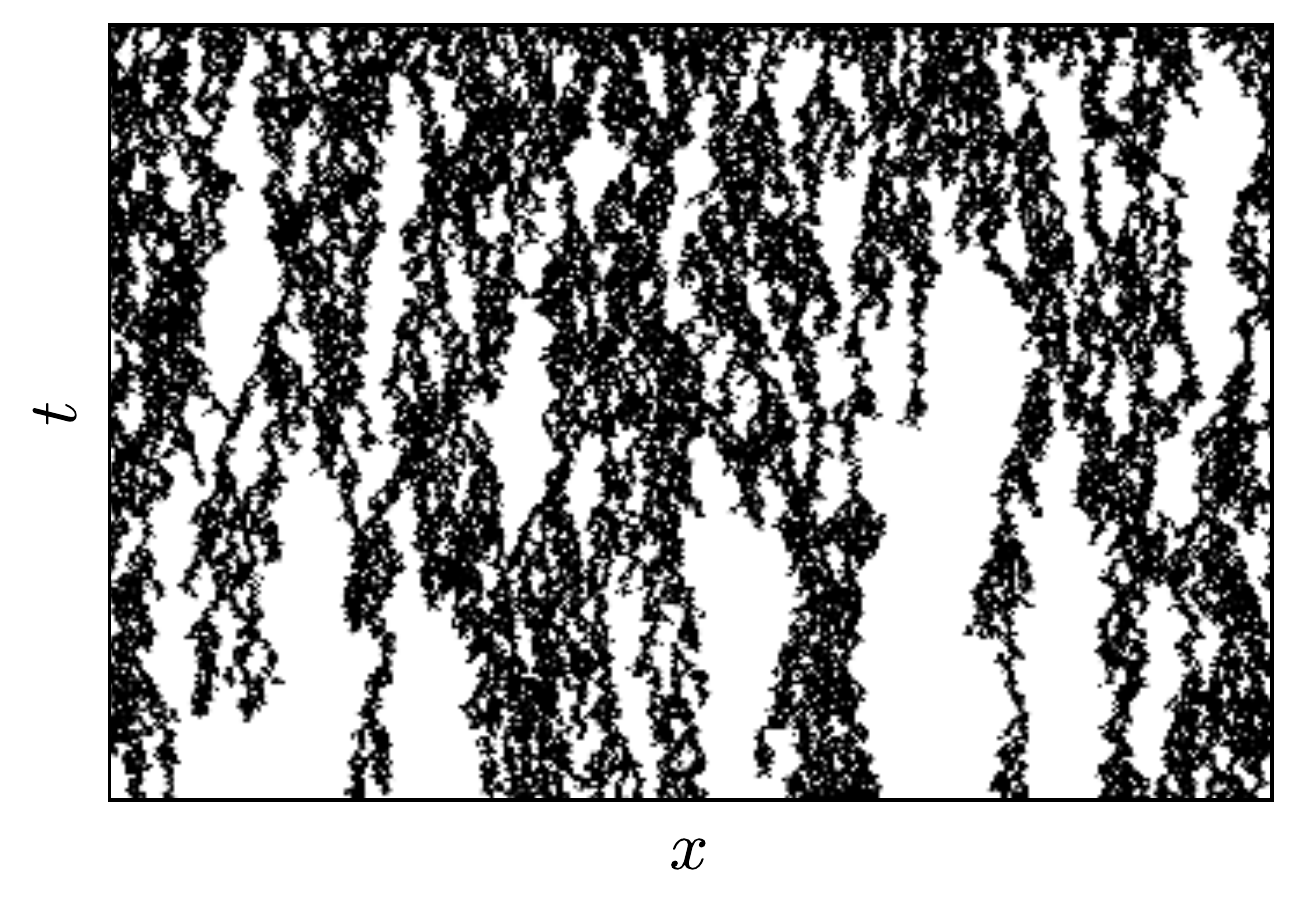}
  \includegraphics[width=0.49\linewidth]{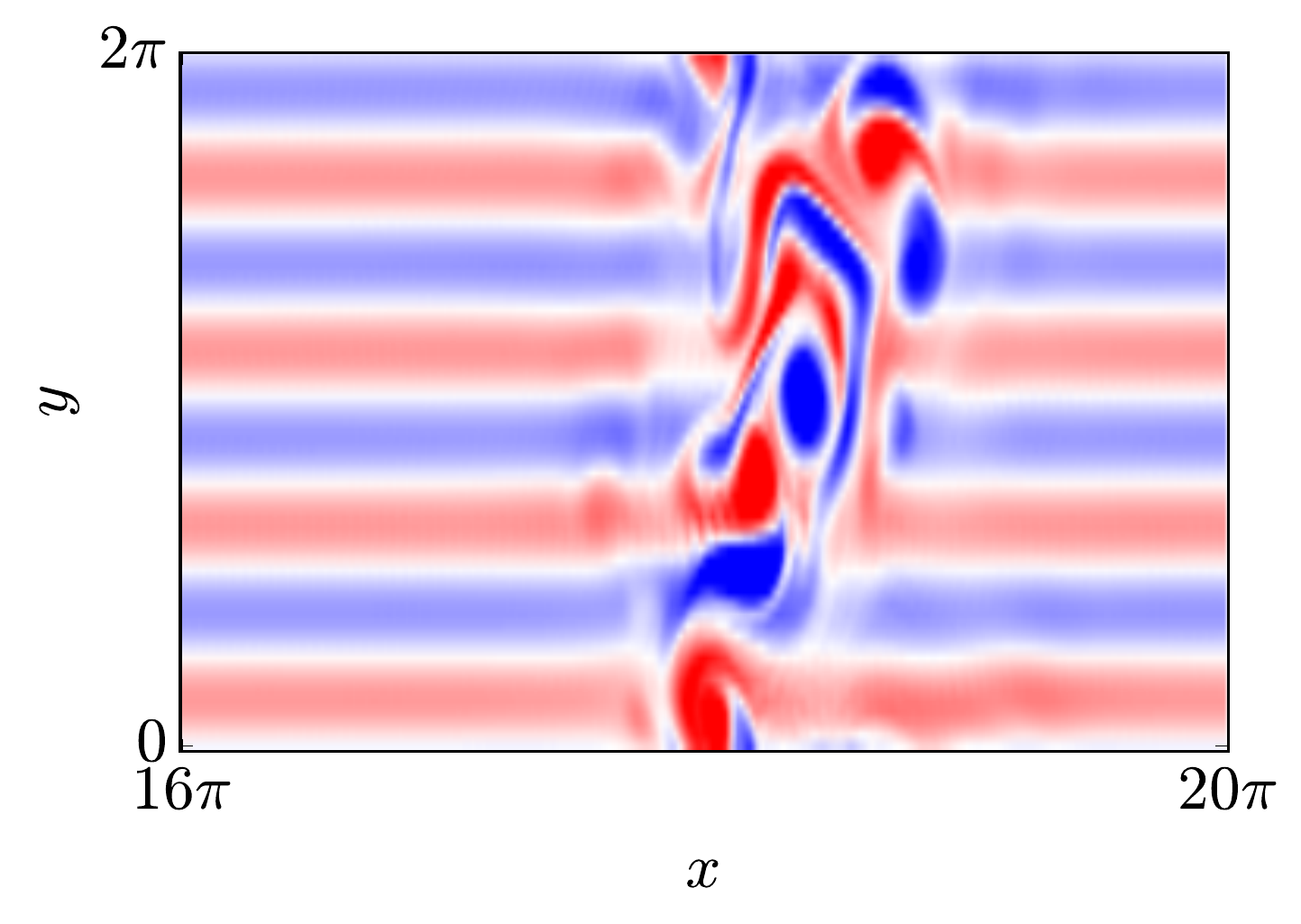}
  \caption{\label{fig:evol} Visualization of the flow for (Re $, $Re$\gamma,\
 U_y,\ n,\ \alpha)=(242,\ 30,\ 0.5,\ 4,\ 1/256)$. (a) Typical time evolution of turbulent region(black). White region denotes 
  laminar region. (b) Snapshot of a part of the whole domain, $x \in
 [16\pi,20\pi]$, using vorticity $\omega$.  The whole domain is $x \in [0,512\pi]$.}
\end{figure}

In this letter, we report observations of subcritical laminar-turbulence transition 
belonging to DP universality class in pure 2D flow as shown in FIG.\ref{fig:evol}.
We also discuss the origin of linear stability of the laminar solution 
and the locality of dynamics in this Navier-Stokes model.

We focus on two-dimensional (2D) Kolmogorov flow with the linear drag force
in a doubly periodic box
$(x,y) \in [0,2\pi/\alpha]\times[0,2\pi]$.
The velocity field $\bm{u}=(u_x,u_y)$ is governed by 2D Navier-Stokes equation 
with a steady sinusoidal force in non-dimensionalized form as follows:
\begin{align}
 \partial_t\bm{u}+\bm{u}\cdot\bm{\nabla}\bm{u}&=-\bm{\nabla}P-\gamma(\bm{u}-U_y\bm{\hat{y}})\nonumber\\
 &+\frac{1}{\rm Re}\nabla^2\bm{u}+\sin(ny)\bm{\hat{x}}\label{eq:NS}, \\
  \bm{\nabla}\cdot\bm{u}&=0,
 \end{align}
 where  $\alpha$, Re, $\gamma$, $n$, $\hat{\bm{x}}$ and $\hat{\bm{y}}$
  denote the aspect ratio of the
 rectangle domain, Reynolds number, the coefficient of drag force,
 the wave number of the external force, the unit vectors in $x$ and $y$, respectively.
 The pressure function $P$ is doubly periodic.
 The velocity averaged over the box  and
 the flow rate in $y$-direction are denoted by $\bm U$ and $U_y$ which
 are defined as follows:
\begin{align}
  \bm{U}&=\frac{\alpha}{4\pi^2}\int dx dy \bm{u}=(0,U_y).
 \end{align}

 Note that since the $y$ dependence of the external force breaks Galilean invariance in $y$, the flow rate $U_y$ is a control parameter of the system.
 To nondimensionalize the governing equation
 we use $L_y/2\pi$ for the characteristic length scale where $L_y$ is
 the length of the box in $y$-direction.
 For the work rate of the force to be $\mathcal{O}(1)$,
 the characteristic time is set to  $\sqrt{L_y/2\pi\chi}$ where $\chi$ is
 the amplitude of the sinusoidal force.
 The linear drag force is typically originated from weak but
 inevitable effects of material walls. 
 For the scales longer than the non-dimensional length
$L_{\rm drg}\sim(\gamma {\rm Re})^{-1/2}$, the drag term dominates
  the viscous term in Eq.(\ref{eq:NS}). 
  We expect that the drag force kills large scale structures of the
  system size formed by the energy transport toward large scales
  as occurring in 2D inverse-cascade turbulence\cite{Kraichnan1967,Kraichnan1980,Laurie2014}.

Our DNS solves the governing equation for the vorticity
$\omega=\partial_yu_x-\partial_xu_y$ with the
pseudo-spectral method for spatial
discretization using the two-thirds rule for dealiasing and
the 2nd order Runge-Kutta (Heun) method for time evolution.
The time and spatial resolutions used for DNSs are $2\times10^{-3}$ and
64 points per $2\pi$, respectively. 

 When a subcritical laminar-turbulence transition  is realized in 2DKF,
 the laminar state should be linearly stable at least.
 First, we check the stability of the following laminar state 
 which is the stationary solution with global $x$-translational symmetry
 and denoted by $\bm{u}_{\rm lam}(U_y)$:
\begin{align}
  \bm{u}_{\rm lam}(U_y)&=\frac{\rm Re}{D}\sin(ny+\theta)\hat{\bm{x}}+U_y\hat{\bm{y}}.\label{eq:laminar}
\end{align}
Here, $\tan\theta={\rm Re}U_y n/(n^2+\gamma)$ and $D^2=(n^2+\gamma{\rm Re})^2+(n{\rm Re}U_y)^2$.
This one-parameter form of Eq.(\ref{eq:laminar}) can be applied
approximately  for a part of the whole system and
this locally-embedded laminar profile can take
a certain value of the local flow rate, i.e., $\beta$ which is not
necessarily equal to the system flow rate $U_y$.
Hereafter, we call $\bm{u}_{\rm lam}(U_y)$ 
as the global laminar (GL)
state and  $\bm{u}_{\rm lam}(\beta)$ as a local laminar state, respectively.

For $\gamma=0$, we can estimate the  critical Reynolds number
for the linearly unstable GL state Re$_{\rm l}$ under the assumption 
that the GL state is unstable for a long-wave disturbance in $x$-direction.

The vorticity equation for $\gamma=0$ is rewritten in terms of the doubly periodic stream function $\Psi$ as follows:
\begin{align}
 \partial_t \nabla^2 \Psi  &-\{\Psi,\nabla^2\Psi \}_{xy}\nonumber\\
 &+U_y\partial_y\nabla^2\Psi -\frac{1}{\rm Re}(\nabla^2)^2\Psi=n\cos(ny),\label{eq:PSIeq}
\end{align}
where $\{f,g\}_{xy}=\partial_xf\partial_yg-\partial_yf\partial_xg$.

We search a stationary solution that is the GL state with small long-wave modes
 by the perturbation expansion:
\begin{align}
  \Psi(x,y)=\Psi_{\rm lam}(y)+\sum_{i=0}\epsilon^i\Psi_i(\epsilon x,y), \label{eq:PSI}
\end{align}
 where $\bm{u}_{\rm lam}(U_y)=(\partial_y\Psi_{\rm lam}(y), U_y)$
 and $\epsilon$ is the small parameter.

 Since $\partial_y\Psi_0=0$ in the limit of $\epsilon\rightarrow0$,  we set $\Psi_0=A(\epsilon x)$.
 Executing the perturbation expansion with Eq.(\ref{eq:PSIeq}) yields
\begin{align}
  {\cal L}_y\Psi_i&=F_i(\Psi_{\rm lam},\Psi_0,\cdots,\Psi_i-1), \label{eq:eqp}\\
  {\cal L}_y&=\frac{1}{\rm Re}\partial_y^4-U_y\partial_y^3,
\end{align}
for the i-th order of $\epsilon$.
We assume that there exists a non-trivial solution of
${\cal L}_y\Psi^0_{\rm e}=0$, then $\Psi_{\rm e}^0$ should be a function
independent of $y$.
A necessary condition that each order of Eq.$(\ref{eq:eqp})$ has
a solution is that the inhomogeneous term of Eq.$(\ref{eq:eqp})$
is orthogonal to  $\Psi_{\rm e}^0$ as follows:
\begin{align}
  \int dy F_i(\Psi_{\rm lam},\Psi_0,\ldots,\Psi_i-1)=0 \label{eq:sol0}
\end{align}
at any $X=\epsilon x$.
This condition (\ref{eq:sol0}) is rewritten as
\begin{align}
  \int dy( \{\Psi,\partial_X^2\Psi\}_{Xy}+\epsilon\frac{1}{\rm Re}\partial_X^4\Psi) =0 \label{eq:sol1}
\end{align}
for each order of $\epsilon$.
At the zeroth order Eq.(\ref{eq:sol1}) is automatically satisfied and 
\begin{align}
  \Psi_{\rm lam}&=-\frac{\rm Re}{Dn}\cos(ny+\delta).
\end{align}
At the first order 
\begin{align}
\partial_X^3 \int dy( - (\partial_y\Psi_L)\Psi_1+\frac{1}{\rm Re}\partial_XA)=0.\label{eq:solvab}
\end{align}
We assume that $\Psi_1=({\rm Re}/D)^2\partial_XA\sin(ny+2\delta)$, then 
we obtain the following condition for Reynolds number Re:
\begin{align}
  (1-2n^2U_y^4){\rm Re}^4-4U_y^2n^4{\rm Re}^2-2n^6=0. \label{eq:R4}
\end{align}
Therefore, the condition for $\Psi_1$ to exist is that Re of
Eq.($\ref{eq:R4}$) is positive real.
When $|U_y|<U_{y}^{\rm c}(n)=1/(\sqrt[4]{2}\sqrt{n})$,
there exists the following unique real solution of Eq.($\ref{eq:R4}$):
\begin{align}
  {\rm Re}={\rm Re}_{\rm l}=\sqrt{\frac{2U_y^2n^4+n^3\sqrt{2}}{1-2n^2U_y^4}}. \label{eq:REC}
\end{align}
This solution is a likely candidate for the neutral
stability curve of the GL state:  Re$_{\rm c}=$Re$_{\rm l}(n,U_y)$.
In fact, In the case $U_y=0$, we obtain the same result as the previous works\cite{Sivashinsky1985,Lucas2014c}.
Equation(\ref{eq:REC}) shows the flow rate  stabilizes the  GL state
which becomes linearly stable at any Re when $|U_y|>U_{y}^{\rm c}(n)=(2n^2)^{(-1/4)}$.
We expect this lower bound, i.e., $U_{y}^{\rm c}(n)$ at $\gamma=0$ also
comes into effect at $\gamma>0$ for most cases.

The stabilization of the GL state for large $U_y$ can be understood as follows.
The amplitude of the laminar flow has an upper bound proportional to $U_y^{-1}$.
Thus, for large $U_y$ the amplitude of the GL state is too small to be unstable.
In the following numerical experiments at the subcritical regime,
we choose $U_y > U_y^{\rm c}$ in order for laminar states to be stable.

\begin{figure}[h]
  \includegraphics[width=0.49\linewidth]{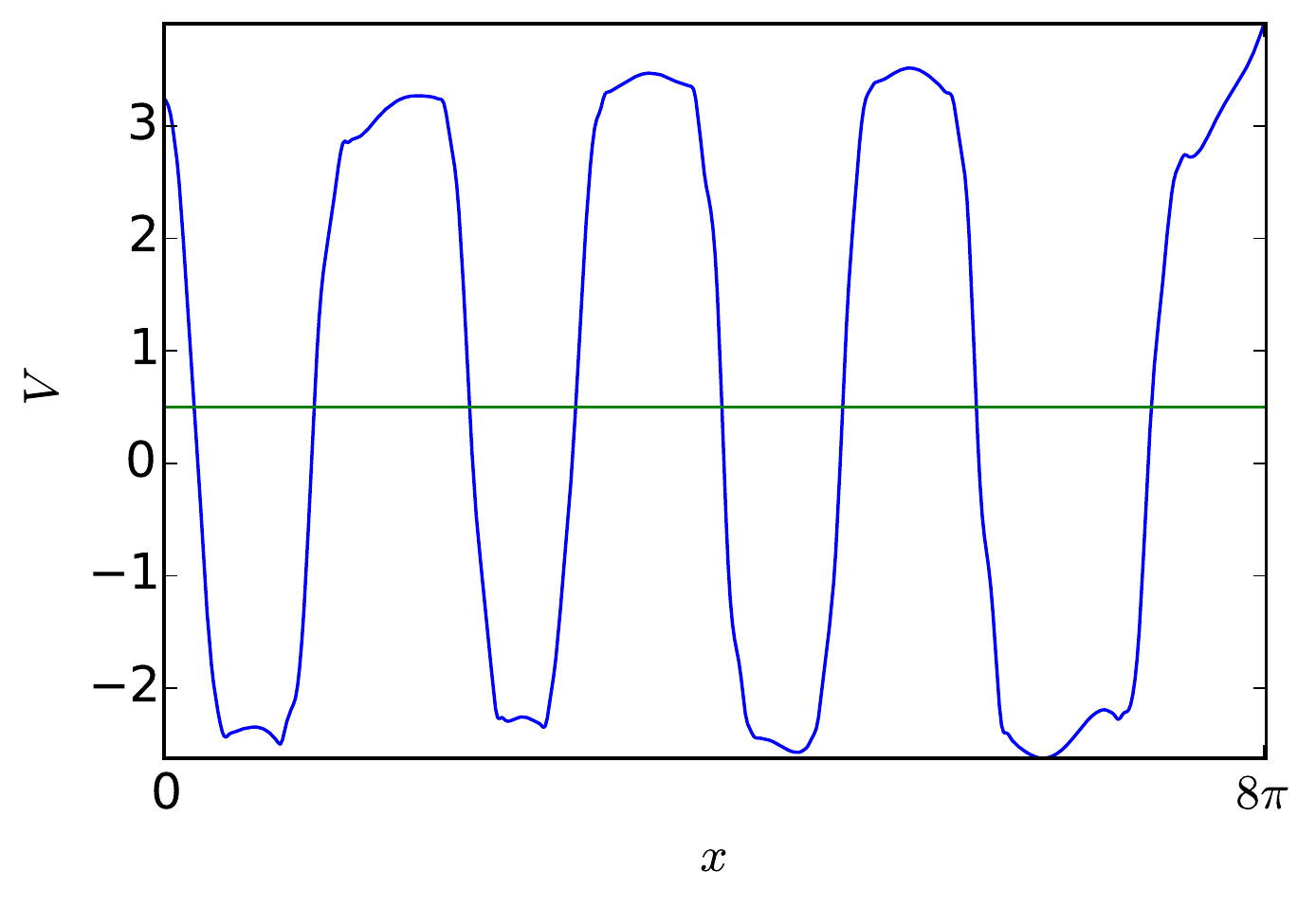}
  \includegraphics[width=0.49\linewidth]{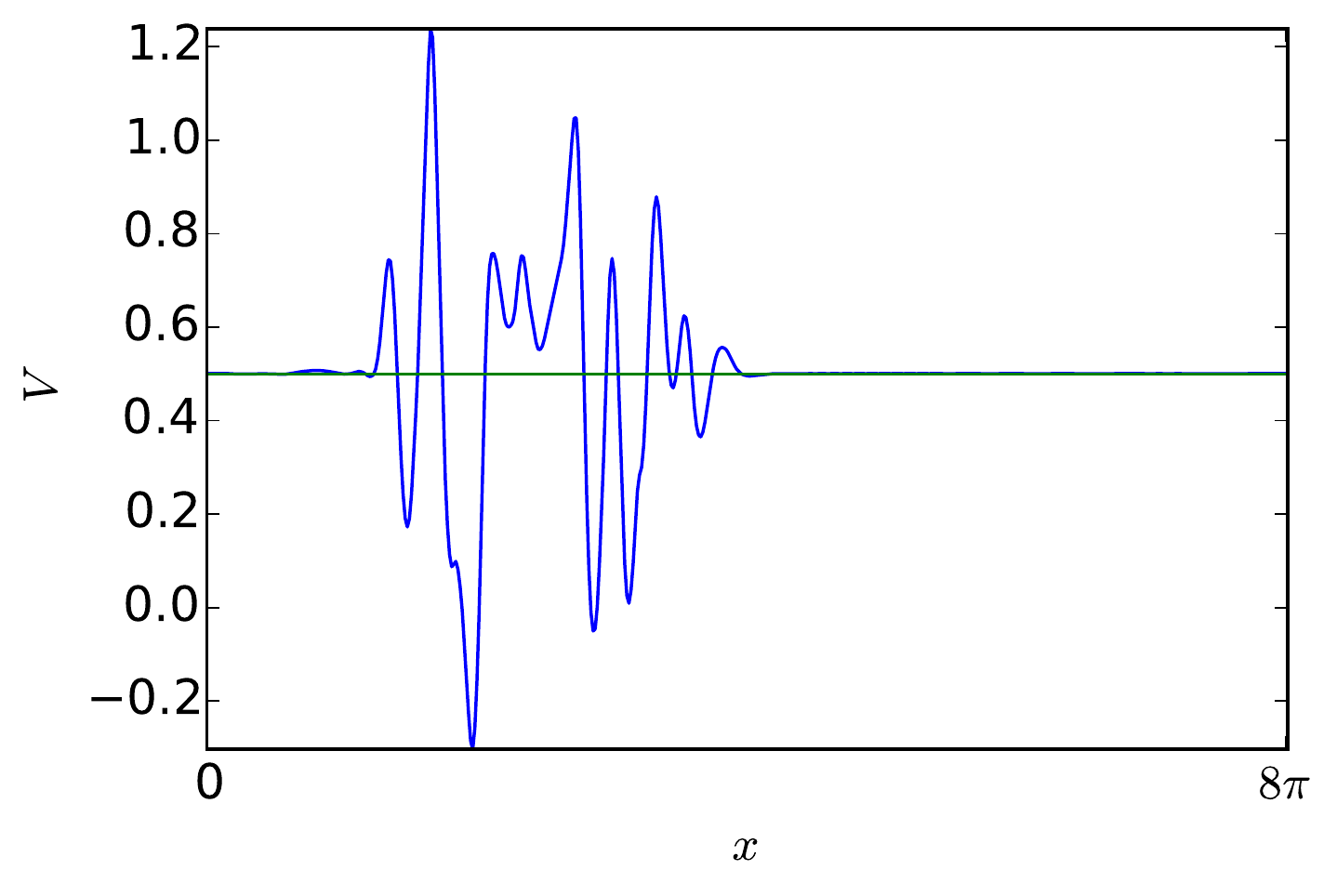}
 \caption{\label{fig:VCV} Average flow over $y$ direction, $V(x) =
 \int_0^{2\pi} dyu_y(x,y)/2\pi$ for Re$=240$. 
  (a)$\gamma{\rm Re}=0$, (b)$\gamma{\rm Re}=30$.  Straight line ($V=0.5$) denotes the GL state.}
\end{figure}

Next, we check the effect of the drag forcing to coherent structures of
the  flow numerically.
We focus on the distribution of velocity in y-direction,
since $U_y$ is a conserved and effective indicator of redistribution
of momentum.
When $\gamma$Re is small, the whole state is composed of kink-antikink
 arrays connecting between adjacent local laminar states with
 $\beta\neq U_y$ as shown in FIG.\ref{fig:VCV}a.
 This state is reminiscent of Cahn-Hilliard like flow emerged
 at the weak  nonlinear regime for $U_y=0$ and $\gamma=0$ 
 where the GL state is linearly unstable for large Re\cite{Lucas2014c,Hiruta2015}.
 When $\gamma$Re is large, the most part of the whole domain are filled
 with the GL state and strong vorticity regions emerge
 spatially-intermittently as shown in FIGs.\ref{fig:evol}b and \ref{fig:VCV}b.
 These kinds of structures are strongly connected with the locality of
 the dynamics of SLT.

To simplify the situation,
we assume the whole domain is separated in $x$-direction into two types
of regions in terms of $V(x)=\int_0^{2\pi} dyu_y(x,y)/2\pi$.
First type is the region which is a laminar state with $V \sim V_0$
and occupies  most of the whole domain.
The other type is SLT whose
characteristic width is $L_{\rm SLT}$.
There are $N$ turbulent states and  $V$ averaged over the width of the
i-th state is denoted by $V_i$.
Then, integrating $V(x)$ over the whole domain, we obtain the relationship 
among $V_0$ and $V_i$: 
\begin{align}
  \sum_{i=1}^N\frac{V_{i}-V_0}{N}\sim(U_y-V_0)/\rho \label{eq:qeff}
\end{align}
where we assume $NL_{\rm SLT}/L$ ($L=2\pi/\alpha$) is approximated by the turbulence fraction denoted by $\rho$.
Since the left hand side means the average distance (norm-like) from the
globally laminar state and the right hand side depends  on the inverse of $\rho$,
this relationship means that the dynamics of SLT directly depends on the number of SLT, $N$ or the turbulence
fraction, $\rho$ when $U_y \neq V_0$ for $\gamma$Re is small.
Therefore, SLTs feel difficulty to decay at finite turbulence fraction $\rho$ and 
the whole state could not decay to the global laminar state.
Therefore, the condition $V_0\sim U_y$ should be satisfied 
for a subcritical transition occurred in 2DKF to belong to DP universality class.
This condition holds for large $\gamma$Re in our system.

 We numerically conduct quenching experiments and confirm the occurrence
 of the DP class transition under the conditions
 that  $n=4$, $U_y=0.5>U_y^{\rm c}(4)\sim0.42$ and $\gamma$Re$=30$ in  
 large domains of  $1/\alpha=64$ and $256$.

 Initial velocity fields are obtained from sustaining turbulence
 consisting of many SLTs at Re$=1000$,
 then Re is reduced to the target values of Re $\in[230,250]$.
 Typically, SLTs make their own replicas
 and decay to local laminar states as randomly and independently.

 As the numerical definition of the local turbulence state by binalization,
 we adopt  the following local distance in 1D form to the local laminar state
 denoted by $L_{\rm loc}(x)$:
\begin{align}
  L_{\rm loc}(x)=\frac{1}{2\pi \delta x}\int_0^{2\pi}dy
  \int_{x-\frac{\delta_x}{2}}^{x+\frac{\delta_x}{2}}dx'(\omega(x')-\omega_{\rm lam})^2,
 \end{align}
 where $\omega_{\rm lam}$ denotes the vorticity of GL state.
  Then, turbulent states are defined by satisfying the
  condition that  $\{x| L_{\rm loc}(x)>{\rm const}\}$.
  The threshold constant is not sensitive to the results so that 
  we choose $\delta x=\pi/16$ which is comparable to the characteristic
  width of the typical SLT,  $L_{\rm SLT}$.
  FIG.\ref{fig:evol}a shows typical visualization of turbulent region 
  in this definition.

\begin{figure}[h]
  \includegraphics[width=0.49\linewidth]{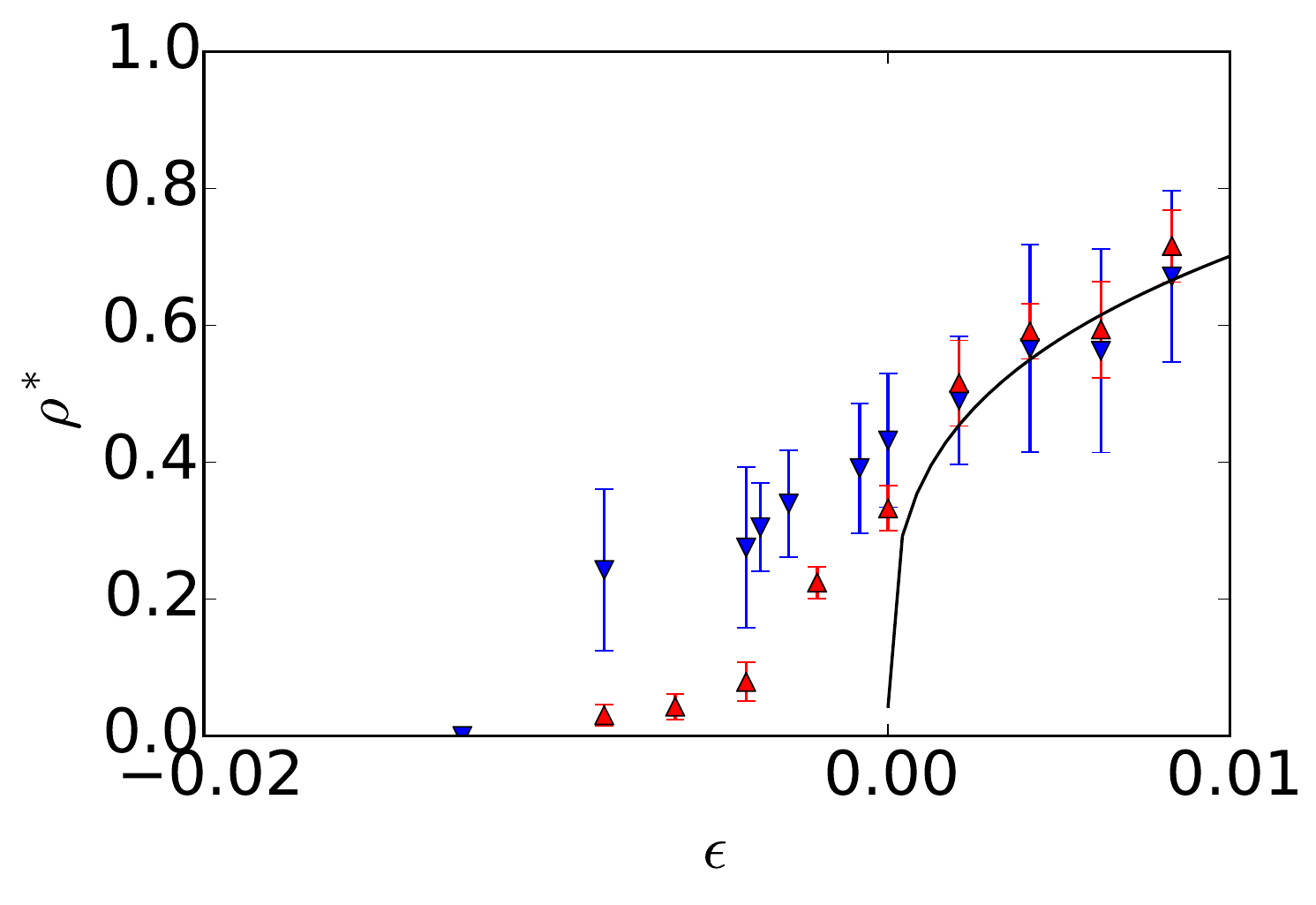}
  \includegraphics[width=0.49\linewidth]{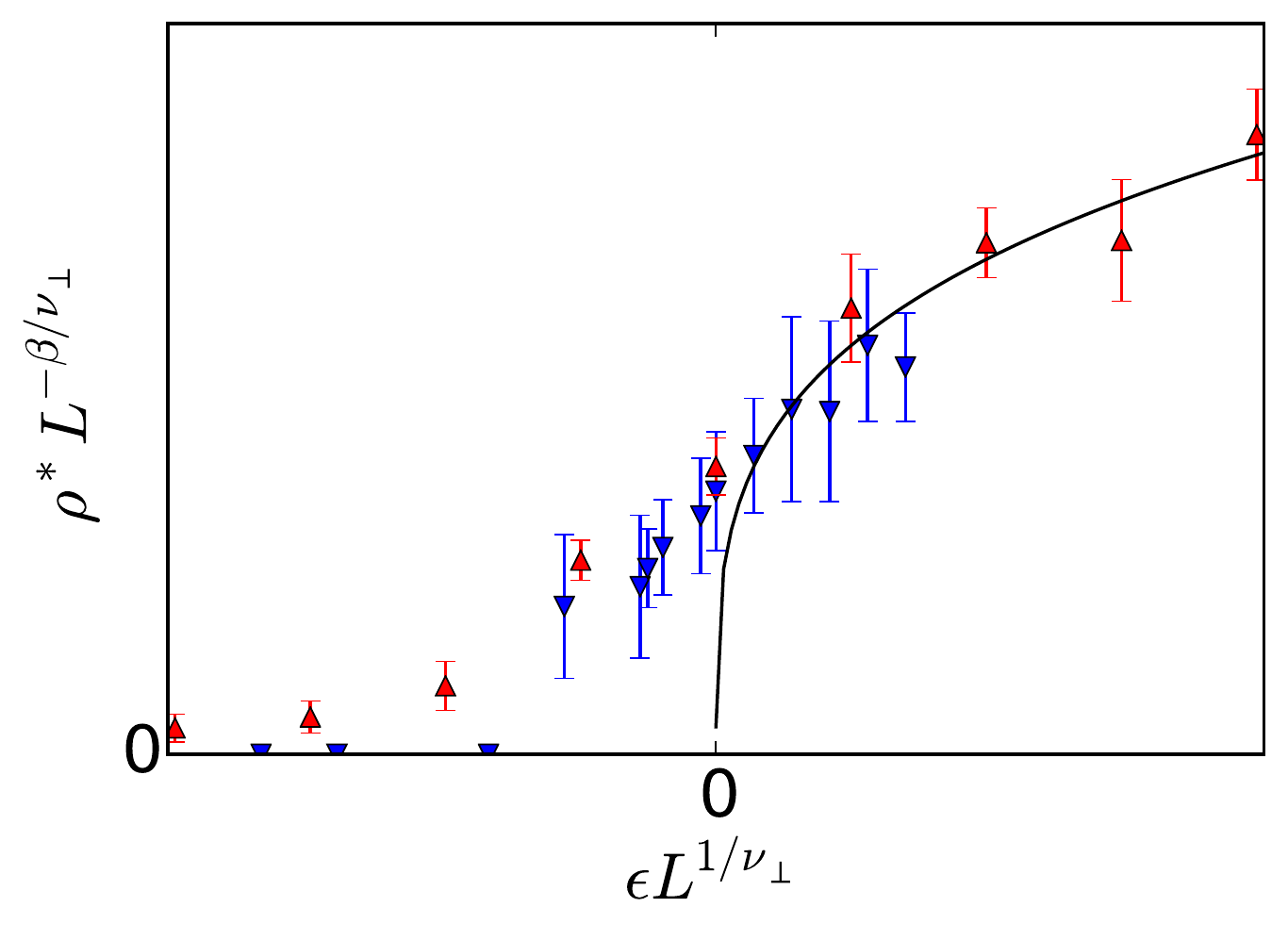}
   \caption{\label{fig:rho}(a) Turbulence fraction and (b) universal scaling function with finite scaling.  
   Lines indicate prediction of (1+1)D DP. Red upper triangles denote  $\alpha=1/256$ and blue lower triangles denote $\alpha=1/64$.
    }
 \end{figure}

\begin{figure}[h]
  \includegraphics[width=0.49\linewidth]{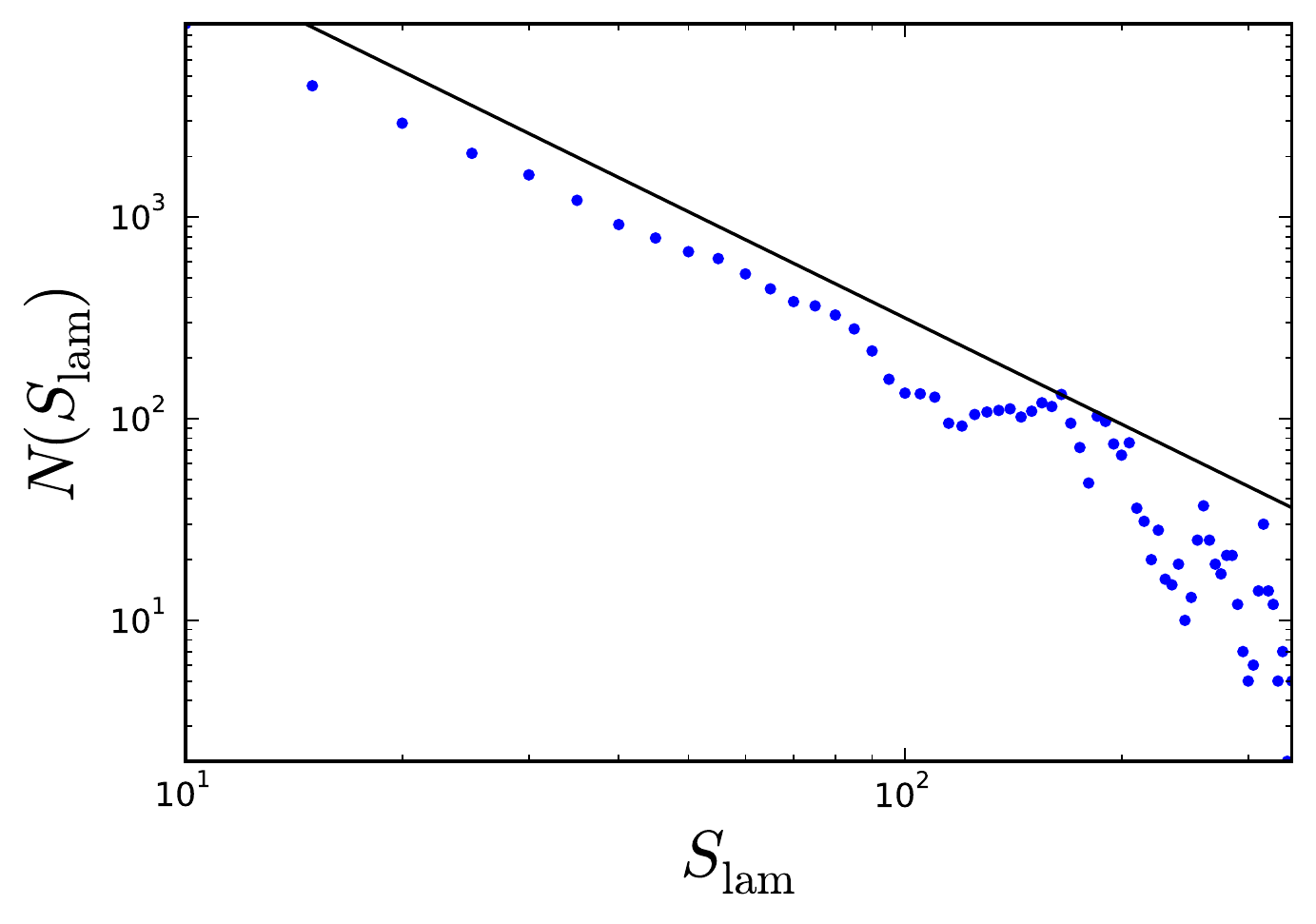}
  \includegraphics[width=0.49\linewidth]{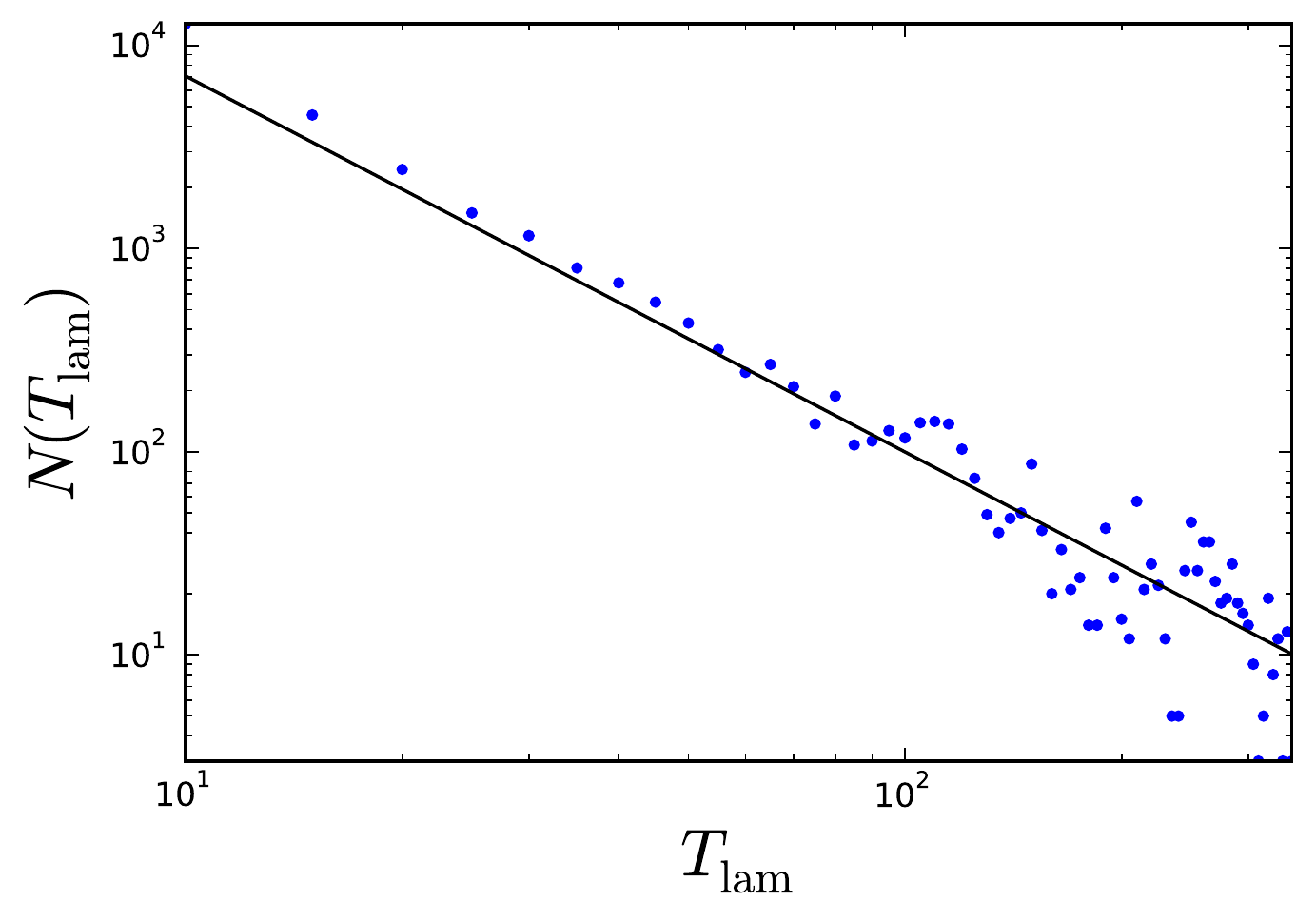}
  \caption{\label{fig:spte}Histogram of laminar gaps (a) in spatial direction (b) in temporal direction. Straight lines indicate prediction of (1+1)D DP.
  }
\end{figure}

To confirm some aspects of the DP critical phenomenon observed in our 2D system,
we estimate the turbulence fraction and  the distributions of laminar
gaps both in space and time. Their dependence on Re is also examined.
Based on the estimated critical Reynolds number Re$_{c}=241$,
we confirm that the three independent critical indices
($\beta$, $\mu_{\perp}$, $\mu_{\parallel}$) are consistent with
the predictions by the (1+1)D DP universality class as  shown in
FIGs.\ref{fig:rho} and \ref{fig:spte}.
Note that these indices are estimated by the following scaling
relations with $\epsilon={\rm Re}/{\rm Re}_c-1$: 
\begin{align}
  \rho^{*}(\epsilon)&\sim \epsilon^{\beta},\label{eq:1+1rho}\\
  N(S_{\rm lam})&\sim S_{\rm lam}^{-\mu_{\perp}},\\
  N(T_{\rm lam})&\sim T_{\rm lam}^{-\mu_{\parallel}},
\end{align}
 where $\rho^{*}$ is 
 $\rho(t)$ for large $t$,
 and $N(S_{\rm lam})$ and  $N(T_{\rm lam})$ are  the histograms of the laminar
 gaps between turbulent region in spatial direction $S_{\rm lam}$ and temporal direction $T_{\rm lam}$, respectively.
 For (1+1)D DP class, the values of the indexes we used are
 $\beta=0.276$, $\nu_{\perp}=1.097$,
 $\nu_{\parallel}=1.734$, $\mu_{\perp}=1.748$ and
 $\mu_{\parallel}=1.841$ which were estimated in ref.\cite{JENSEN1996}.

Turbulence fraction $\rho^{*}$ shown in FIG.\ref{fig:rho}a is consistent with a continuous transition
of (1+1)D DP especially for $\epsilon>0$. Relatively larger error bars are originated from the divergence of long time correlation at the critical point.
For a smaller domain $\alpha=1/64$, $\rho(t)$ quickly decays to zero for $\epsilon<-0.01$.
This is because large fluctuations cause accidentally transitions
 to the absorbing state.
 To estimate a finite size effect, we conducted a finite size scaling
 estimation by considering spatial correlation. 
As shown in FIG.\ref{fig:rho}b, function $L^{\beta/\nu_{\perp}}\rho^{*}(\epsilon/L^{1/\nu_{\perp}})$ 
does not depend on the system size $L$ and scales as $\epsilon^\beta$.
Then, we can expect  that as  $L\rightarrow\infty$  $\rho^{*}(\epsilon)$ 
approaches the scaling (\ref{eq:1+1rho})
 since the $\epsilon^\beta$ part in $\epsilon>0$ remains invariant but
  $\rho$ in $\epsilon<0$ goes to zero.

Each of the two critical exponents $\mu_{\perp}$, $\mu_{\parallel}$ of the GL
state gaps is  connected with the fractal box-counting dimension in each direction.
Both of the histograms in FIG.\ref{fig:spte} show
 consistency with the DP prediction.
 The slope for the spatial gaps in FIG.\ref{fig:spte}a is slightly
 less steeper as discussed in ref.\cite{Chantry2017a}

\begin{figure}[h]
  \includegraphics[width=0.49\linewidth]{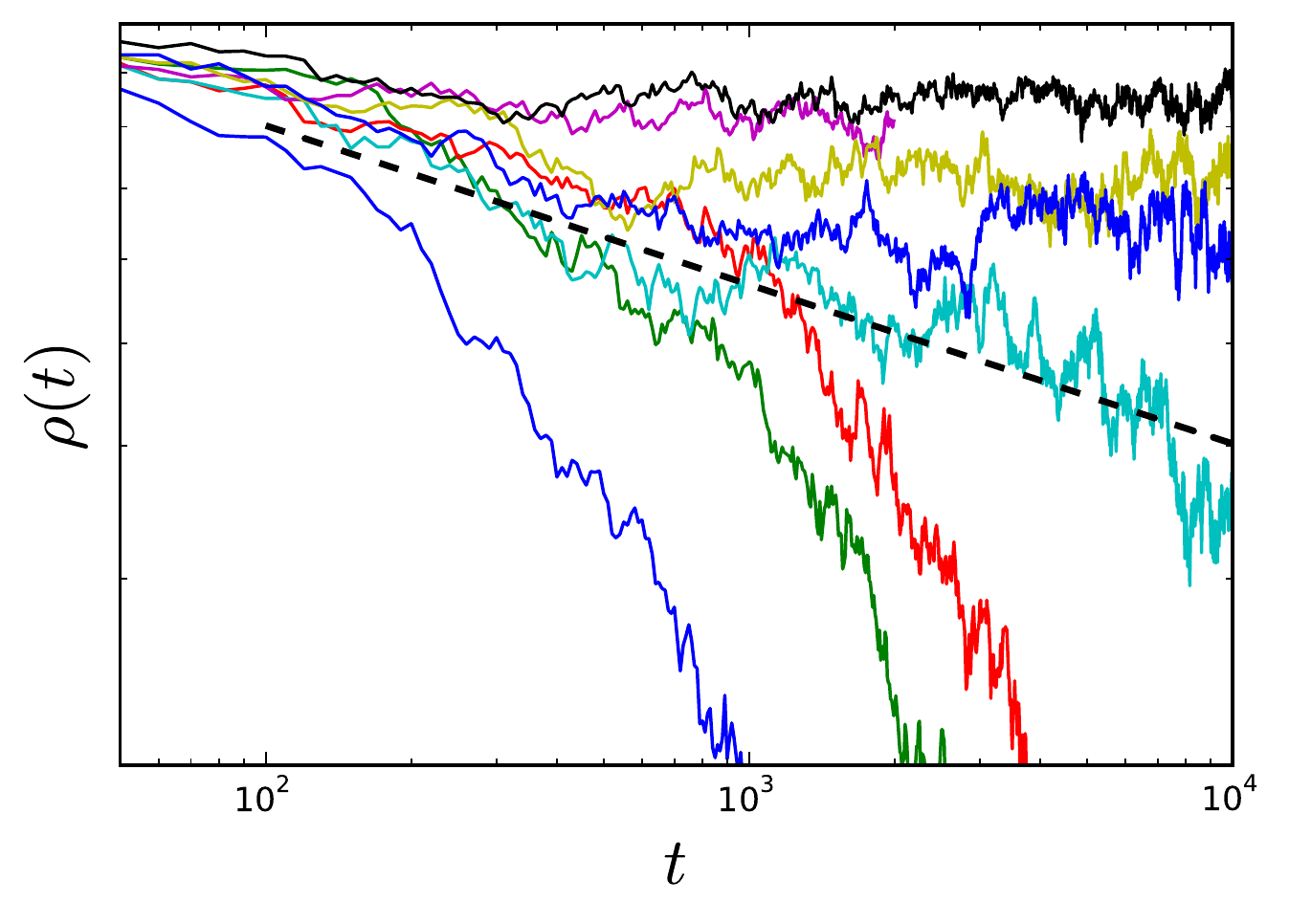}
  \includegraphics[width=0.49\linewidth]{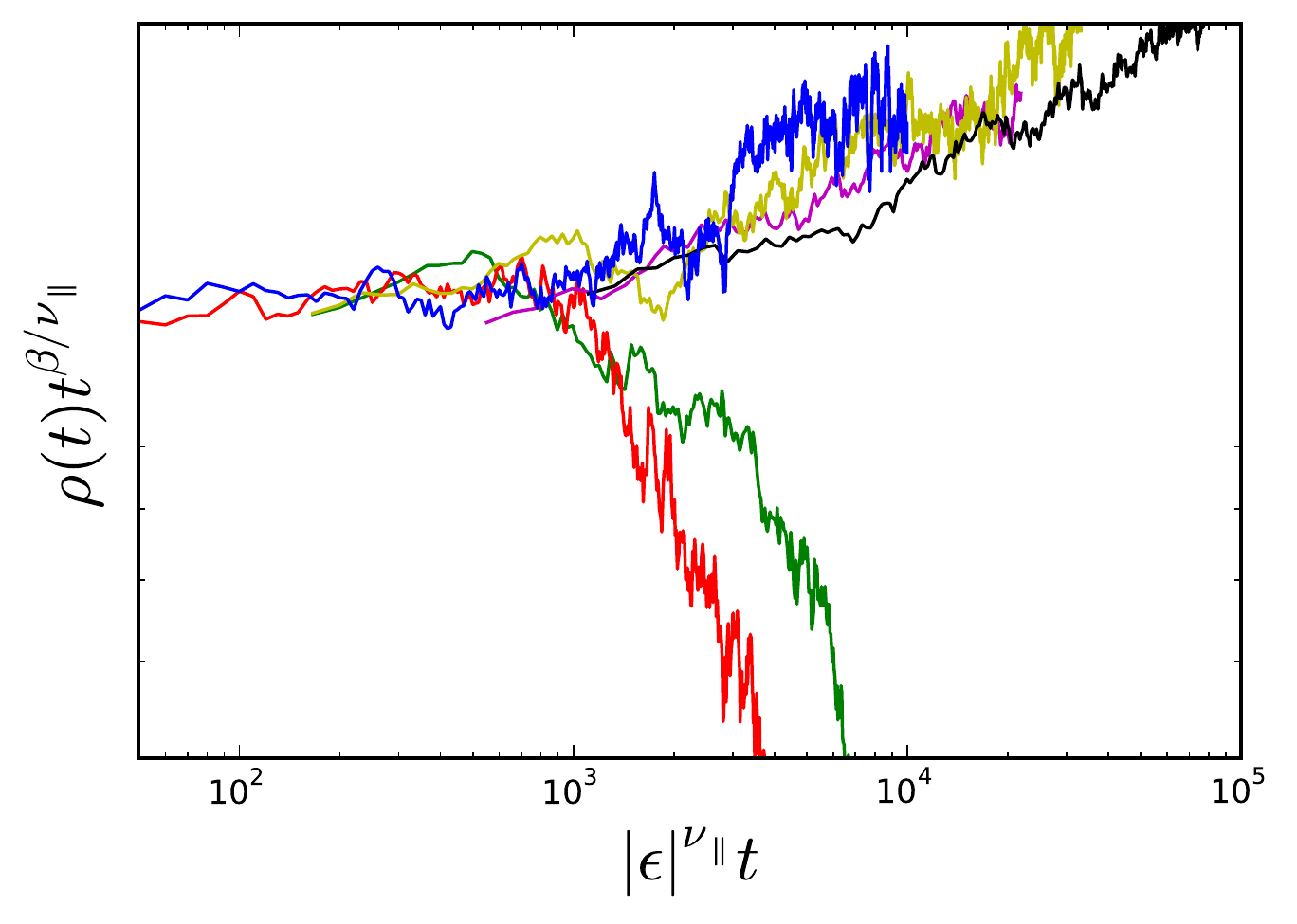}
  \caption{\label{fig:univ} (a)Decaying temporal turbulence fraction $\rho(t)$.
 Hashed line indicate prediction of (1+1)D DP. Re $=235,239,240,241,242,243,245,247$ from the lower left to the upper right.
 (b) Estimation of universal function by pre-multiplied curve. All curve tend to collapse to one line.
  }
\end{figure}

 As shown in FIG.\ref{fig:univ}, 
 by the scaling with the DP critical exponents,
 the decays of the turbulence fraction for quenching processes
 for a wide range of Re values collapse approximately to a single
 universal function,
 $F(t)=\rho(t\epsilon^{-\nu_{\parallel}})t^{\beta/\nu_{\parallel}}$.
 All these results support that the subcritical transition observed in
 2DKF belongs to (1+1)D DP universality class.

In summery, laminar-turbulence transition belonging to DP universality class,
which is commonly seen in wall-bounded flows with large domains, 
 also occurs in a 2D flow with periodic boundary conditions.
This result explicitly shows the concrete form(s) of self sustained
process(SSP) for near-wall dynamics is not necessarily essential
for subcritical transition observed in wall-bounded flows. 
Moreover, we can also conclude the DP class transition realized
for subcritical turbulence transition is not universal. This is
supported by counterexamples observed for Kolmogorov flow with $\gamma=0$.

The essential characteristics of DP, the existence of the absorbing state and 
the locality, are related in our system to the flow rate $U_y$ and
the drag forcing $-\gamma\bm{u}$.
This parameter dependency is difficult to be identified for wall-bounded flows 
since they typically have only one independent parameter Re.

We now consider what plays the roles of $U_y$ and $\gamma$
in wall-bounded flows or more general flows.
As discussed above, the drag forcing can be interpreted as a frictional
force by the walls.
From the governing equation ($\ref{eq:NS}$), $U_y$ can be interpreted as
an advection speed of a weak disturbance in the frame of zero flow rate:
the forcing term is $\sin(n(y-U_y t))$ in this frame.
This advection effect disturbs an efficient supply of kinetic energy for disturbances
to grow.
This interpretation is consistent with the fact 
that the tails of SLT do not develop for large $U_y$ flows\cite{Hiruta2015,Hiruta2017a}.
These kinds of effects can be seen in wall-bounded flow but not be controlled.
An advantage of our system is that the important effects can be
easily separated in terms of a multidimensional parameter space.
Thus, we expect that even more general flows share the characteristics
of our simple system and that more detailed researches help us to
grasp fundamental aspects of sustenance of turbulence.

To confirm the above conjecture,  we need to estimate or describe
theoretically and quantitatively the interaction or correlation among localised turbulent structures.
These estimation and description are important not only for the consideration of minimum elements of DP transition
but also for analyses of collective behaviors of localised turbulent
structures which appear as large scale structures such as turbulence stripes 
in the wall-bounded turbulence\cite{Coles1965,Prigent2002,Duguet2010,Tuckerman2011,Ishida2016}.

We have dealt with laminar-turbulence transition observed in 2DKF which
is governed by the 2D Navier-Stokes equation. This governing equation
must  be the simplest in the  physically-supported spatio-temporal
evolution equations which can describe the transition of DP universality class.
In this sense, we expect that we can study the relationship between
governing law and universality class of nonequilibrium phase transition
from more general contexts.\\

This work was supported by JSPS KAKENHI Grant Number JP18J13334,
Grant-in-Aid for JSPS Research Fellow;
the Research Institute for Mathematical Sciences, a Joint
Usage/Research Center located in Kyoto University.

\bibliography{main}

\end{document}